\documentclass[showpacs,aps,pre,twocolumn,amsmath,amssymb,epsfig]{revtex4-1}
\usepackage{epsfig}

\newcommand     {\beq}[1]         { \begin{equation} #1 \end{equation} }
\newcommand     {\beqa}[1]        { \begin{eqnarray} #1 \end{eqnarray} }

\begin{document}

\title{Blending stiffness and strength disorder can stabilize fracture}

\author{Ehud D. Karpas}
\affiliation{Weizmann Institute of Science, Rehovot, 7610001, Israel}
\author{Ferenc Kun}\email{ferenc.kun@science.unideb.hu}
\affiliation{Department of Theoretical Physics, University of Debrecen, 
P.O.Box:5, H-4010 Debrecen, Hungary}

\date{\today}

\begin{abstract} 
Quasi-brittle behavior where macroscopic failure is preceded by stable
damaging and intensive cracking activity is a desired feature of materials 
because it makes fracture predictable. Based on a fiber bundle model with global 
load sharing we show that blending strength and stiffness disorder of material
elements leads to the stabilization of fracture, i.e.\ samples which are brittle
when one source of disorder is present, become quasi-brittle as a consequence 
of blending. We derive a condition of quasi-brittle behavior in terms of the 
joint distribution of the two sources of disorder.
Breaking bursts have a power law size distribution of exponent $5/2$ 
without any crossover to a lower exponent when the amount
of disorder is gradually decreased. The results have practical relevance 
for the design of materials to increase the safety of constructions.

\end{abstract}
\pacs{64.60.av,46.50.+a,81.40.Np}
\maketitle

\section{Introduction}
Disorder is an inherent property of practically all materials, both natural and
man made. The heterogeneity occurring on different length scales plays
a crucial role in the mechanical response and fracture behavior of materials. 
On the one hand, the presence of flaws, voids, and grain boundaries reduces
the fracture strength, on the other hand, however, they improve
the damage tolerance of materials which has a high practical 
importance for construction components \cite{herrmann_statistical_1990,
alava_statistical_2006,book_chakrabarti_2015}. 
Low disorder leads to brittle behavior where fracture occurs at a critical load
in a catastrophic manner without any precursors. From a practical point of view 
quasi-brittle behavior is desired, which is typical for materials with a higher amount
of disorder. The fracture process of quasi-brittle materials is composed of a large
number of intermittent steps of cracking giving rise to the emergence of crackling noise
\cite{PhysRevLett.110.165506,PhysRevE.85.016116}.
Analyzing the statistics and dynamics of crackling noise, methods can be worked out 
to forecast the imminent global failure \cite{vasseur_heterogeneity:_2015}. 
Controlling the amount of disorder to
enhance the quasi-brittle behavior of materials has a high practical relevance.

The theoretical investigation
of the fracture of heterogeneous materials is mainly 
based on discrete stochastic models such as Fiber Bundle (FBM) 
\cite{kun_damage_2000,hidalgo_universality_2008,hidalgo_avalanche_2009,
pradhan_failure_2010},
fuse model \cite{nukala_percolation_2004} and 
Discrete Element (DEM) approaches 
\cite{daddetta_application_2002,carmona_fragmentation_2008,
ergenzinger_discrete_2010} with Monte Carlo and molecular 
dynamics simulation techniques.
A common basic assumption of such modeling approaches is that the
heterogeneity of materials can be fully captured by introducing
disorder for the local fracture strength of cohesive elements of the model.  
However, recent experimental and theoretical investigations led to the 
surprising 
conclusion that the heterogeneity of the local stiffness 
can improve the fracture toughness of materials 
\cite{okumura_toughness_2004,urabe_fracture_2010}. 

Beyond artificially made (tailored)
materials, structural heterogeneity of local stiffness is an important feature
of biological materials, as well 
\cite{okumura_why_2001,nukala_continuous_2005,layton_equal_2006,
barthelat_biomimetics_2007}. 
One of the most known example of such biological composite materials is nacre 
which exhibits extraordinary mechanical properties 
compared to its constituent materials. These features can be partly attributed 
to the interplay of the local variation of stiffness and strength 
\cite{okumura_why_2001,nukala_continuous_2005}.

In the present paper we consider this problem in the framework of fiber bundle 
models (FBM)
by introducing two sources of disorder, i.e.\ both the stiffness and strength of 
fibers
are random variables. Assuming global load sharing after failure events we obtain 
a 
generic analytical description of the mechanical response of the system on the 
macro-level,
and investigate the microscopic process of failure by computer simulations. 
We show that blending stiffness and strength disorder results in stabilization
of fracture even if the system with a single source of disorder had a perfectly 
brittle response.

\section{Fiber bundle with two random fields}
In the model we consider a parallel set of $N$ fibers which is loaded 
along the fibers' direction. Fibers have a perfectly brittle
response, i.e.\ they exhibit a linearly elastic behavior with a Young
modulus $E$ and fail when the load $\sigma$ on them exceeds a threshold value 
$\sigma_{th}$. 
In order to capture the local variation of stiffness and strength of materials, 
it is a crucial element of the model that each fiber is characterized by two 
random variables: 
the Young modulus of fibers $E$ takes values in the interval $E_{-} \leq E 
\leq E_{+}$ 
according to the probability distribution $f(E)$, while the breaking threshold
$\sigma_{th}$ is sampled with the probability density $g(\sigma_{th})$ over the
interval $\sigma_{-} \leq \sigma_{th} \leq \sigma_{+}$. 
In the present study the two random fields are assumed to be independent,
i.e.\ no correlation is considered between strength and stiffness.
Hence, in a finite bundle of $N$ fibers two independent random values $\sigma_{th}^i$
and $E_i$ are assigned to each fiber $i=1, \ldots, N$.
When fibers fail during the stress controlled loading of the bundle the load of 
broken 
fibers has to be overtaken by the remaining intact ones. For simplicity, we
assume infinite range of load sharing which can be ensured by loading the 
bundle between 
two perfectly rigid platens. It has the consequence that the strain 
$\varepsilon$ of fibers is always 
the same, however, due to the randomness of the Young modulus $E$ their local 
load $\sigma_i$ is a fluctuating quantity $\sigma_i = E_i\varepsilon$, 
where $i=1,\ldots , N$. 

In the classical FBM 
\cite{pradhan_failure_2010,hidalgo_avalanche_2009,yoshioka_size_2008} the 
strength 
of fibers $\sigma_{th}$ is the only random variable
which represents the heterogeneity of materials. Since the Young modulus 
is constant $E=const.$,
in the limit of global load sharing of the model fibers keep the same load $E\varepsilon$, 
and hence, break in the increasing order of their breaking threshold 
$\sigma_{th}^i, i=1, \ldots , N$. 
It has the consequence that the macroscopic 
constitutive equation $\sigma_0(\varepsilon)$ can be 
expressed in terms of the cummulative distribution 
of strength thresholds
$G(\sigma_{th})=\int_{\sigma_-}^{\sigma_{th}}g(x)dx$ 
in the form
\beq{
\sigma_0(\varepsilon) = E\varepsilon \left[1-G(E\varepsilon)\right].
\label{eq:srandom}
}
Here the term $\left[1-G(E\varepsilon)\right]$ provides the fraction 
of intact fibers at the strain $\varepsilon$, and $\sigma_0$ denotes 
the external load.

In the opposite limit of the model all fibers have the same breaking threshold
$\sigma_{th}=const.$, however, their Young modulus $E$ is random with $E_i$, $i=1,\ldots , N$
values. The breaking condition
$E_i\varepsilon > \sigma_{th}$ implies that in this case fibers break in the 
decreasing order of their Young moduli $E_i$. Recently, we have 
shown that in this case the constitutive equation of the model 
can be obtained as \cite{0295-5075-95-1-16004}
\beq{
\displaystyle{
\sigma_0(\varepsilon) = \varepsilon \int_{E_{-}}^{\sigma_{th}/\varepsilon} E f(E)dE,}
\label{eq:erandom}
}
where $f(E)$ is the probability density of the Young modulus of fibers.
Equation (\ref{eq:erandom}) expresses that at a given strain $\varepsilon$ those 
fibers are intact in the bundle whose stiffness $E$ falls below $\sigma_{th}/\varepsilon$.

Our present fiber bundle model is a combination of the above two cases allowing 
for randomness both in the stiffness $E$ and strength $\sigma_{th}$ of fibers.
In the presence of two disorder fields $E$ 
and $\sigma_{th}$ the breaking sequence becomes more complex: 
Fibers break when the load on them $E_i\varepsilon$ exceeds the local breaking threshold
$E_i\varepsilon>\sigma_{th}^i$, hence, at a strain $\varepsilon$ those fibers are broken 
for which the condition $\varepsilon>\sigma_{th}^i/E_i$ holds. It can be seen that the breaking
sequence of fibers is controlled by the ratio of their strength and stiffness which defines 
their critical strain of breaking $\varepsilon_{th}^i=\sigma_{th}^i/E_i$, and hence, the 
breaking condition can be formulated as $\varepsilon>\varepsilon_{th}^i$.
It follows that in our model of random stiffness and strength with global load sharing 
fibers break in the increasing order of the local failure strain 
$\varepsilon_{th}^i$ ($i=1,\ldots , N$). 
\begin{figure}%[!h]
  \begin{center}
\epsfig{bbllx=30,bblly=10,bburx=350,bbury=300,file=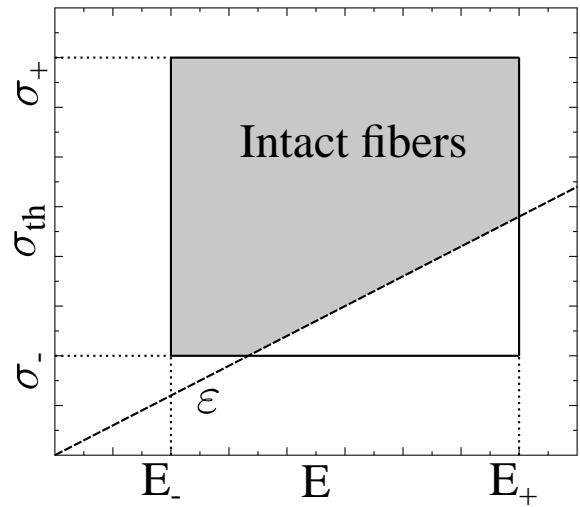, width=8.0cm}
   \caption{Failure plane of fibers for uniformly distributed threshold values.
   Each point with parameter values ($E,\sigma_{th}$) inside the rectangle of side lengths
   $E_+-E_-$ and $\sigma_+-\sigma_-$ represents a fiber of the bundle.
   The equation of the dashed straight line is
$\sigma_{th}=E\varepsilon$ so that fibers above the line fullfill 
the condition $\sigma_{th}>E\varepsilon$.
At a given strain $\varepsilon$ during the loading process those fibers 
are intact (highlighted with gray color) and keep the external load, which fall 
above this line of slope $\varepsilon$. 
The integral in Eq.\ (\ref{const}) has to be performed over the domain of intact fibers.
\label{fig:failure_plan} }
\end{center} 
\end{figure}

Since the stiffness and the failure strength are independent
random variables, the load carried by the intact fibers having Young modulus 
and 
strength in the interval $[E, E+dE]$ and $[\sigma_{th}, \sigma_{th}+d\sigma_{th}]$, 
respectively,
reads as $E\varepsilon f(E)g(\sigma_{th})dEd\sigma_{th}$.
Integrating the contributions of all intact fibers we obtain the generic form
of the constitutive equation
\begin{equation}  %General constitutive equatinon
\displaystyle{
\sigma_0(\varepsilon) = 
\varepsilon \int_{\sigma_{-}}^{\sigma_{+}} 
\int_{E_{-}}^{\sigma_{th}/\varepsilon} 
Ef(E)g(\sigma_{th})dEd\sigma_{th},}
\label{const}
\end{equation}
in terms of the probability density functions $f$ and $g$ of the Young modulus 
and strength of fibers, respectively.
First the integral over $E$ has to be performed, where the upper limit 
$\sigma_{th}/\varepsilon$ of the integral 
captures the effect that at the macroscopic strain $\varepsilon$ only those 
fibers can be intact 
which have a Young modulus below $\sigma_{th}/\varepsilon$ 
(similarly to Eq.\ (\ref{eq:erandom})). 
Then the integral over the strength $\sigma_{th}$ of single 
fibers follows, where $\sigma_{th}$ can take any value in the range
$\sigma_{-}\leq\sigma_{th}\leq \sigma_{+}$.

It can be observed that taking the small strain limit 
$\varepsilon \rightarrow 0$ 
in the constitutive equation Eq.\ (\ref{const}) we restore linear behavior in the form 
$\sigma_0(\varepsilon \rightarrow 0) 
= \varepsilon \left< E \right>$ where $\left< E \right>$
denotes the average Young modulus of fibers $\left< E \right>=\int_{E_{-}}^{E_{+}}Ef(E)dE$.
In the large strain limit the macroscopic stress goes to zero 
$\sigma_0(\varepsilon \rightarrow \infty) \rightarrow 0$ 
since there are no intact fibers left.
It is important to note that setting the probability distribution of the Young 
modulus or breaking threshold to a Dirac delta function $f(E)=\delta(E-E_0)$ and 
$g(\sigma_{th})= \delta(\sigma_{th}-\sigma_{th}^0)$, 
the constitutive equation Eq.\ (\ref{const}) of our model
recovers the FBM equations Eqs.\ (\ref{eq:srandom}) and (\ref{eq:erandom}) 
with only one source of disorder for failure strength
\cite{pradhan_failure_2010,hidalgo_avalanche_2009,yoshioka_size_2008} and for 
the Young modulus \cite{0295-5075-95-1-16004}, respectively.

In the following we investigate the breaking process of our FBM both on the 
macro and micro scales. 
In order to clarify the effect of blending strength and stiffness disorder
we focus on systems which exhibit perfectly brittle failure if only one source 
of disorder is present. 

\section{Uniformly distributed stiffness and strength}
The details of the macroscopic response of the system and of the breaking 
process of fibers can easily be determined analytically when both the Young modulus $E$
and the strength $\sigma_{th}$ of fibers are uniformly distributed with
the probability densities 
\beq{
\displaystyle{ f(E) = \frac{1}{E_{+}-E_{-}}, \qquad \mbox{and} 
\qquad g(\sigma_{th}) = \frac{1}{\sigma_{+}-\sigma_{-}}.}
\label{eq:unif}
} 
For brevity, we define the notation 
$\displaystyle{ B\equiv f(E) g(\sigma_{th}) = 1/[(E_+-E_-)(\sigma_+ - \sigma_-)]}$, 
and the strains where the first and last fibers break are denoted by
$\varepsilon_{min} \equiv
\sigma_-/E_+$, $\varepsilon_{max} \equiv \sigma_+/E_-$, respectively. 
In addition, we introduce
$\varepsilon_1 \equiv \sigma_-/E_-, \
\varepsilon_2 \equiv \sigma_+/E_+$ and point out that $\varepsilon_1$ can be 
smaller or larger than $\varepsilon_2$ depending on the parameters of the density 
functions Eq.\ (\ref{eq:unif}).

\subsection{Macroscopic response}
\label{sec:macro}
The macroscopic constitutive curve $\sigma_0(\varepsilon)$ of the bundle can be
obtained by inserting the above probability densities Eq.\ (\ref{eq:unif}) 
into the generic form Eq.\ (\ref{const}). 
Figure \ref{fig:failure_plan} illustrates the failure plane $(\sigma_{th}, E)$ 
where the breaking thresholds $\varepsilon_{th}$ of fibers can take values. 
When the strain $\varepsilon$ is reached
during the loading process those fibers remain intact and keep the load which 
fall above the 
straight line of slope $\varepsilon$ (it is highlighted with gray color in the 
figure).
\begin{figure}%[!h]
  \begin{center}
\epsfig{bbllx=10,bblly=10,bburx=350,bbury=305,file=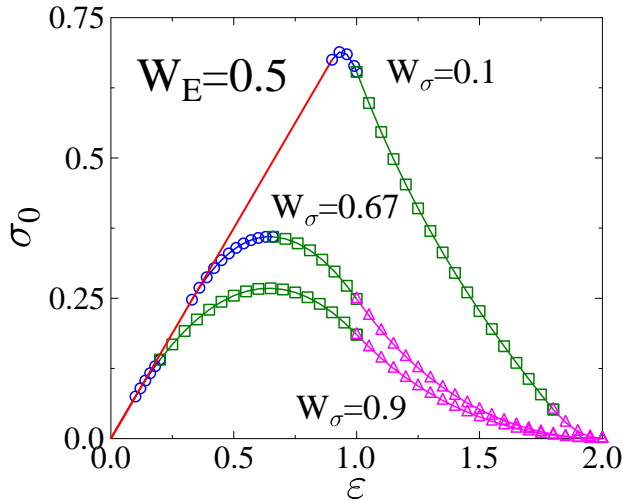,
width=8.0cm}
   \caption{(Color online) Macroscopic response of a system where both 
   sources of disorder are  uniformly distributed with the parameter  $W_E=0.5$ 
 for three different values of $W_{\sigma}$. 
 Different symbols and colors are used to highlight the regimes corresponding
 to different terms of the integral expression Eq.\ (\ref{macrouniform}): bold line (red),
 circle (blue), square (green), and triangle (magenta) stand for the 
contributions of the first, second, third, and fourth terms of Eq.\ (\ref{macrouniform}).
 \label{macroscopic_uni}}
\end{center} 
\end{figure}
Assuming the case $\varepsilon_1<\varepsilon_2$ 
the integrals of Eq.\ (\ref{const}) can be carried out in a piecewise manner as
\begin{widetext}
\begin{equation}
\displaystyle{\sigma_0(\varepsilon) =} \left\{  
\begin{array}{llll} \displaystyle
& \displaystyle{0.5 (E_+ +E_-) \varepsilon}, & \quad 
&\displaystyle{\varepsilon<\varepsilon_{min}}; \\[2.5mm]
& \displaystyle{\frac{1}{6B} [-2 E_+^3 \varepsilon^2  - 
\frac{\sigma_-^3}{\varepsilon}  + 
  3(E_+^2 \sigma_+ + E_-^2 \sigma_- - E_-^2\sigma_+)\varepsilon]} , & \quad & 
\displaystyle{\varepsilon_{min}<
  \varepsilon < \varepsilon_1}; \\[2.5mm]
& \displaystyle{\frac{1}{6B} [-2 (E_+^3 - E_-^3) \varepsilon^2 + 3\sigma_+(E_+^2 
- E_-^2)\varepsilon]}, & \quad & 
  \displaystyle{\varepsilon_1< \varepsilon < \varepsilon_2}; \\[2.5mm]
& \displaystyle{\frac{1}{6B} [2 E_-^3 \varepsilon^2 - 3 \sigma_+ E_-^2 
\varepsilon +
  \frac{\sigma_+^3}{\varepsilon}]} , & \quad & \displaystyle{\varepsilon_2 < 
\varepsilon < \varepsilon_{max}}; \\[3mm]
& 0, & \quad & \displaystyle{\varepsilon_{max} < \varepsilon }.            
\label {macrouniform}
\end{array}
\right.
\end{equation}
\end{widetext}
If $\varepsilon_2<\varepsilon_1$ the macroscopic behavior is the same except for 
the
interval $\varepsilon_1<\varepsilon<\varepsilon_2$ (the third interval of 
Eq.\ (\ref{macrouniform})) which is replaced by
\begin{equation} 
\sigma_0(\varepsilon) =
\frac{1}{6B} [-2 (\sigma_+^3 - \sigma_-^3) \frac{1}{\varepsilon}  -
3(\sigma_+-\sigma_-)E_-^2 \varepsilon].
\end{equation}

In this case we must also exchange $\varepsilon_1$ and $\varepsilon_2$ everywhere in the 
limits of
the intervals.
In order to quantify the amount of disorder in the system, without loss of generality,
from here on end 
we fix the upper limits $E_+=1$ and $\sigma_+=1$ and control the disorder by the width 
of the distributions $W_E$ and $W_{\sigma}$ such that $W_E \equiv E_+ - E_-$ and
$W_{\sigma} \equiv \sigma_+ - \sigma_-$. 
The macroscopic response $\sigma_0(\varepsilon)$ of the fiber bundle 
is presented in Fig.\ \ref{macroscopic_uni} for several values of $W_{\sigma}$ 
keeping the width $W_E=0.5$ fixed. 
It can be observed that for strains $\varepsilon<\varepsilon_{\rm min}$, where
no fiber breaking occurs, the system exhibits a perfectly linear response with
an effective Young modulus equal to the average value $\left<E\right>=(E_++E_-)/2$ of E. 
Above $\varepsilon_{\rm min}$ non-linearity emerges as the 
consequence of gradual breaking of fibers indicating the quasi-brittle 
behavior of the bundle. Note that the decreasing regime of 
the constitutive curves in Fig.\ \ref{macroscopic_uni} 
can only be realized under strain controlled loading conditions. Subjecting 
the system to an increasing external load catastrophic collapse occurs when the 
peak of $\sigma_0(\varepsilon)$ is surpassed. The value $\sigma_0^c$ of the peak 
stress
defines the fracture strength of the bundle, while the peak position 
$\varepsilon_c$ 
provides the critical strain. It can be observed  in Fig.\ 
\ref{macroscopic_uni} 
that the extension of the non-linear regime preceding macroscopic failure, 
and hence, the degree of brittleness, strongly depends on the amount of disorder.

\subsection{Fraction of broken fibers}
In order to quantify the degree of brittleness of the system 
we determined the fraction of fibers $P_b$ which break before the collapse 
at $\sigma_0^c$ under stress controlled loading.
Perfectly brittle behavior is characterized by the value $P_b=0$, 
since in this case the breaking of the first fiber gives 
rise to catastrophic failure of the system.
Starting from the constitutive equation Eq.\ (\ref{macrouniform}) 
we can find the critical strain $\varepsilon_c$, and then $P_b$ can be obtained
from the disorder distribution as 
\begin{equation}
\displaystyle{P_b = 1 - \int_{\sigma_-}^{\sigma_+} \int_{E_{-}}^{\sigma_{th} / 
\varepsilon_c}  f(E) g(\sigma_{th})dE d\sigma_{th}.}
\label{Pb}
\end{equation}

\begin{figure}%[!h] 
  \begin{center}
\epsfig{bbllx=0,bblly=0,bburx=450,bbury=360,file=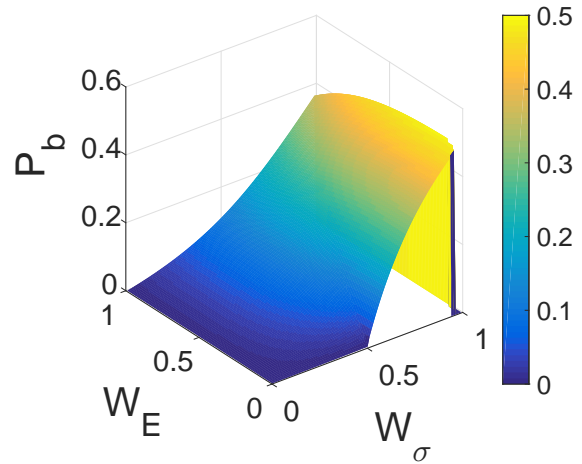, width=8.0cm}
  \caption{(Color online) The fraction of fibers $P_b$ which break before the peak of the 
constitutive curve is reached during stress controlled loading of the bundle.
The surface of $P_b(W_E,W_{\sigma})$ was obtained by numerically evaluating the 
analytical 
expression Eq.\ (\ref{Pb}). $P_b$ has a finite value
everywhere on the $W_E$ - $W_{\sigma}$ plane except on the two axis. \label{fig_Pb}}
\end{center} 
\end{figure}

It has been shown analytically in the classical 
fiber bundle model, i.e.\ in the absence of stiffness disorder, 
that the system has a perfectly brittle
response for narrow distributions $W_{\sigma} \leq 0.5$ of fibers' strength 
\cite{pradhan_crossover_2005-1,raischel_local_2006}. Recently, we have
demonstrated that in the opposite limit when the uniformly distributed 
random stiffness is the only source of disorder, the macroscopic response
of the bundle is perfectly brittle at any value of $W_E$ 
\cite{0295-5075-95-1-16004}. 
We evaluated the integral of Eq.\ (\ref{Pb}) by numerical means varying the 
amount of disorder over the entire range $0\leq W_E$, $W_{\sigma}\leq 1$. 
It can be seen in Fig.\ \ref{fig_Pb} that $P_b$ obtains a finite value $P_b>0$ 
everywhere in the $W_E$ - $W_{\sigma}$ plane except for the $W_E$ axis 
where $W_{\sigma}=0$ holds, and in the range $W_{\sigma}\leq 1/2$ on 
the $W_{\sigma}$ axis where $W_{E}=0$ holds.
The result has the astonishing consequence that whenever there is disorder 
present 
both in local strength and stiffness of material elements the macroscopic 
response 
of the system is quasi-brittle, i.e.\ a finite fraction of 
fibers breaks before the catastrophic failure of the bundle,
so that the constitutive curve of the system 
$\sigma_0(\varepsilon)$ is never perfectly linear up to the maximum.  

For the case of $W_E=0$ the integral of Eq.\ (\ref{Pb}) can be carried out
analytically, which yields for the breaking fraction
\beq{
P_b(W_E=0, W_{\sigma}) = \left\{
\begin{array}{lll}
& 0 & \displaystyle{W_{\sigma}<0.5,} \\[1.5mm]
& \displaystyle{1-\frac{1}{2W_{\sigma}},} &  \displaystyle{0.5\leq W_{\sigma}\leq 1}.
\end{array}
\right.
\label{eq:pbanal}
}

This result shows that in the absence of
stiffness disorder $W_E=0$ a transition occurs at $W_{\sigma}=1/2$ 
between a perfectly brittle $P_b=0$ and a quasi-brittle behavior $P_b>0$.
The numerical results demonstrate that for any finite amount of stiffness 
disorder $W_E>0$ the transition disappears since always a finite fraction 
of fibers breaks before failure $P_b>0$.
It follows that blending stiffness and strength disorder 
can stabilize the system in the sense that
no catastrophic collapse can occur without precursors.

\section{Condition of stability}
In the previous section it has been shown using the cumulative quantity $P_b$ 
that mixing stiffness and strength disorder results in stability of the system in the sense 
that immediate catastrophic failure at the time of first breaking is avoided.
In the following we analyze the transition from the perfectly brittle to 
quasi-brittle behavior by focusing on the microscopic dynamics of the failure process. 

We can formulate a criterion for the stability of the fracture process in terms 
of the disorder distributions based on the idea that the system is perfectly 
brittle if the first fiber breaking induces a catastrophic avalanche. 
At the breaking of the first fiber with the threshold value 
$\varepsilon_{th}^{min}=\sigma_-/E_+$
the load on the bundle is $\left<E\right> \varepsilon_{th}^{min}= \left<E\right>\sigma_{-}/E_{+}$.
After the breaking event the new Young modulus can be approximated as 
$\left<E\right>' \approx \left<E\right> - E_{+}/N$, which gives rise to a
higher strain $\varepsilon'$ of the bundle 
\beq{
\varepsilon' = \frac{\sigma_{-}}{E_{+}}
\left[\frac{\left<E\right>}{\left<E\right> - E_{+}/N}\right].
}
Consequently, the strain increment $\Delta 
\varepsilon=\varepsilon'-\varepsilon_{th}^{min}$ 
generated by the breaking event under a fixed load can be cast in the form
\beq{
\Delta \varepsilon = \frac{\sigma_{-}}{N\left<E\right>}.
}
The average number of fiber breakings $a(\varepsilon_{th}^{min})$ induced by 
the first failure reads as
\beq{
a(\varepsilon_{th}^{min})=Nh(\varepsilon_{th}^{min})\Delta 
\varepsilon
=h\left(\frac{\sigma_{-}}{E_{+}}\right)\frac{\sigma_{-}}{\left<E\right>},
}
where $h(\varepsilon_{th})$ denotes the probability distribution of threshold strains 
$\varepsilon_{th}$.
The avalanche induced by the first fiber breaking becomes catastrophic if
$a(\varepsilon_{th}^{min})>1$, which yields the stability condition of our
system
\beq{
h\left(\frac{\sigma_{-}}{E_{+}}\right)\frac{\sigma_{-}}{\left<E\right>}<1.
\label{eq:stability}
} 
Note that the condition is general and can be applied to any disorder distribution.

If there is only stiffness disorder present ($\sigma_{th}^i=\sigma_{th}$, $i=1, \ldots , N$) 
the distribution $h(\varepsilon_{th})$
can be obtained from the stiffness distribution $f(E)$ 
with a simple transformation
\beq{
h(\varepsilon_{th})=f\left(\frac{\sigma_{th}}{\varepsilon_{th}}\right)\frac{\sigma_{
th}}{\varepsilon_{th}^2}.
}
Substituting e.g.\ the uniform distribution Eq.\ (\ref{eq:unif}) we obtain 
the condition $E_{+}/(E_{+}-E_{-})<1/\sqrt{2}$, which can never hold.
It follows that if the stiffness disorder is the only source of heterogeneity
of the system, for uniformly distributed stiffness values the system always
has a perfectly brittle behavior. In Ref.\ \cite{0295-5075-95-1-16004} 
the same result was obtained but in a different way focusing on the shape of the 
constitutive curve.

% When both the stiffness and strength of fibers have disorder the microscopic
% dynamics of the system remains similar, i.e.\ the fibers still break in the 
% increasing order of their failure strain $\varepsilon_{th}^i$, however, now it 
% has
% to be calculated as the ratio of two random variables 
% $\varepsilon_{th}=\sigma_{th}^i/E_i$. 
% In order to understand how the presence of two sources of disorder can stabilize
% the system we focus again on the initiation of the breaking process. In this 
% case
% the first breaking occurs at the thershold strain 
% $\varepsilon_{th}^{min}=\sigma_{th}^{min}/E_{max}$
% which has to be inserted into the generic form of the stability criterion Eq.\ 
% (\ref{eq:stability}).

When both the stiffness and strength of fibers have disorder
the probability distribution of the strain thresholds $h(\varepsilon_{th})$ can 
be calculated
as the convolution of $f(E)$ and $g(\sigma_{th})$ taking the ratio of the two 
random variables $\varepsilon_{th}=\sigma_{th}/E$
\beq{
h(\varepsilon_{th}) = \int_{E_-}^{E_+} Ef(E)g(\varepsilon_{th}E) dE.
\label{eq:convol}
}
We carried out the integration for the specific case when both random variables 
are uniformly distributed and $\varepsilon_1<\varepsilon_2$ holds
\beqa{\label{eq:epsdist}
 h(\varepsilon_{th}) = \left\{
 \begin{array}{lll}
 0 & \displaystyle{ \varepsilon_{th} < \frac{\sigma_{-}}{E_{+}},}  \\[4mm]
 \displaystyle{ \frac{1}{2B}\left[E_{+}^2 - \frac{\sigma_{-}^2}{\varepsilon_{th}^2} \right],}
   &  \displaystyle{\frac{\sigma_{-}}{E_{+}} \leq \varepsilon_{th} < \frac{\sigma_{-}}{E_{-}}, }
 \\[4mm]
 \displaystyle{ \frac{1}{2B}\left[E_{+}^2 - E_{-}^2 \right], }
   &  \displaystyle{\frac{\sigma_{-}}{E_{-}} \leq \varepsilon_{th} < \frac{\sigma_{+}}{E_{+}}, }
 \\[4mm]
 \displaystyle{\frac{1}{2B}\left[\frac{\sigma_{+}^2}{\varepsilon_{th}^2} - E_{-}^2 \right],} 
   &  \displaystyle{ \frac{\sigma_{+}}{E_{+}} \leq \varepsilon_{th} < \frac{\sigma_{+}}{E_{-}}, }
 \\[4mm]
 0 &  \displaystyle{\varepsilon_{th} > \frac{\sigma_{+}}{E_{-}}. }
\end{array}
\right.
}
\begin{figure}%[!h]
  \begin{center}
\epsfig{bbllx=15,bblly=15,bburx=400,bbury=300,file=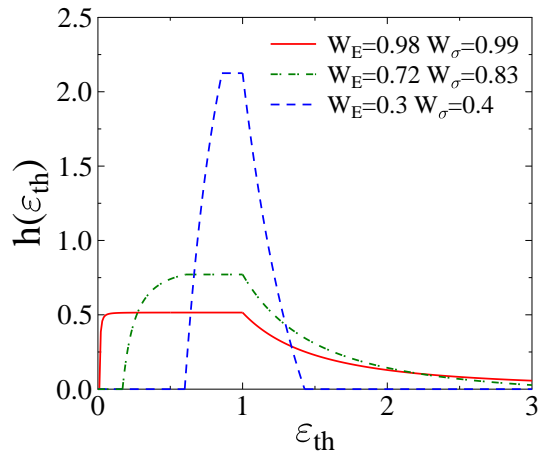, width=8.0cm}
   \caption{(Color online) Probability distribution $h(\varepsilon_{th})$ of the strain thresholds 
   $\varepsilon_{th}$ of fibers calculated from Eq.\ (\ref{eq:epsdist}) for three 
parameter sets. \label{fig:dist_eps_th}}
\end{center} 
\end{figure}
Figure \ref{fig:dist_eps_th} illustrates the distribution $h(\varepsilon_{th})$ 
for three combinations of $W_{E}$ and $W_{\sigma}$. 
 It can be observed that for uniformly distributed 
stiffness and strength the 
distribution of the strain thresholds $h(\varepsilon_{th})$ starts continuously 
from zero for any finite value of $\sigma_{-}$ without any finite jump.
This feature explains why the combination of two perfectly brittle systems leads 
to the emergence of quasi-brittle
behavior where stable cracking precedes macroscopic failure. For the case 
$\sigma_{-}=0$ the distribution
starts with a finite constant, however, the small strain value $\varepsilon\simeq 0$ 
still ensures stability.
For those disorder distributions of stiffness and strength which cover the range 
from 0 to $+\infty$
stability of the blend is again guaranteed by the generic form of the 
distribution Eq.\ (\ref{eq:convol}).

\section{Avalanches of fiber failures}
Under quasi-statically increasing external load $\sigma_0$ when a fiber breaks 
its load gets redistributed over the remaining intact fibers which may induce
further failure events. As a consequence of subsequent load redistribution a 
single
breaking fiber may trigger an entire avalanche of breaking events. 
The randomness of local physical properties and the interaction of fibers 
introduced by the load sharing result in highly complex microscopic 
dynamics of the failure
process \cite{kloster_burst_1997,hidalgo_avalanche_2009}. In the following we
explore the statistics of breaking avalanches of fibers by computer simulations.

\subsection{Computer simulation technique}
First we present the algorithm which allows us to simulate the fracture process 
of large 
bundles. It has been assumed that the bundle is loaded between stiff platens 
which ensures
that the strain of fibers is the same $\varepsilon$. As the external load 
$\sigma_0$ 
is increased the fibers break in the increasing order of their failure strain 
$\varepsilon^i_{th}$,
determined as $\varepsilon^i_{th}=\sigma^i_{th}/E^i$ ($i=1, \ldots , N$) 
which fall in the range 
$\sigma_{-}/E_{+}\leq \varepsilon_{th} \leq \sigma_{+}/E_{-}$.
Computer simulation of the failure process of a finite bundle of $N$ fibers
under stress controlled loading proceeds in the following steps:
\begin{enumerate}
\item Generate random values of the Young modulus $E_i$, and failure strength 
$\sigma_{th}^i$, 
$i=1,\ldots , N$ of fibers according 
      to the desired distributions $f(E)$ and $g(\sigma_{th})$. 
      Independence of the random fields has to be ensured.
\item Determine the failure strains $\varepsilon_{th}^i=\sigma_{th}^i/E_i$, 
$i=1,\ldots , N$,
      and sort them into ascending order.
\item Increase the externally imposed strain $\varepsilon$ up to the smallest 
      threshold $\varepsilon=\varepsilon_{th}^{min}$, and remove the breaking 
      fiber.
      At this instant there is $\sigma_0=\left<E\right>\varepsilon$ load on the 
      system, where the initial value of the average Young modulus of the bundle 
      reads as
\beq{
  \left<E\right> =\frac{1}{N}\sum_{i=1}^{N} E_i.
}
\item After the breaking event the load of the broken fiber gets redistributed 
      over the remaining intact ones keeping the external load $\sigma_0$ constant.
      This load redistribution may induce additional fiber failures and 
      eventually can even trigger an avalanche of breaking events. Note that due 
      to the random stiffness, fibers keep a different amount of load that is why 
      long range interaction is easier to realize 
      through the control of strain. Triggered breakings can be determined 
      in the following way: after the breaking event the overall Young modulus 
      of the bundle has to be updated 
 \beq{
 \left<E\right>' = \frac{1}{N_{int}}\sum_{i\in I} E_i,
%  \left<E\right>' =\frac{1}{N-1}\sum_{i=1}^{N-1} E_i.
\label{eq:update_e}
}
where $I$ denotes the set of intact fibers which has $N_{int}$ elements.

      Since $\left<E\right>'<\left<E\right>$ and the external load is kept 
      constant, the strain of the bundle increases to the new value  
      $\varepsilon'$, which can be obtained as 
\beq{
      \varepsilon' = \left<E\right> \varepsilon/\left<E\right>'.
}
\item Those fibers which have threshold values below the the updated strain 
      $\varepsilon_{th}^j<\varepsilon'$ have to be removed and the algorithm is 
continued
      with step 4. During bursts of breaking one has to take into account in 
Eq.\ (\ref{eq:update_e})
      that more than one fiber may also break in an iteration step.

\item If no more fibers break due to load redistribution, the avalanche ended and 
      the external strain can be 
      increased again to the strain threshold of the next intact fiber in the 
      sorted sequence of $\varepsilon_{th}$.
\end{enumerate}
The efficiency of the algorithm enabled us to simulate bundles of $N=10^7$ 
fibers averaging over 5000 samples at each parameter set with moderate CPU times.

\subsection{Size distribution of bursts}
Using the above algorithm we explored the bursting activity
accompanying the stress controlled loading process of quasi-brittle systems
with two sources of disorder. The avalanches are characterized by their size
$\Delta$, which is defined as the number of fibers breaking in the correlated trail
of the avalanche.
For the classical FBM, where the fiber strength is the only source of disorder,
it has been shown analytically \cite{kloster_burst_1997,pradhan_failure_2010,
hidalgo_avalanche_2009} 
and by computer simulations \cite{hidalgo_avalanche_2009,kun_damage_2000-1}
that for equal load sharing the size distribution of avalanches $p(\Delta)$ 
has a power law behavior
\beq{
p(\Delta)  \sim \Delta^{-\tau}.
\label{eq:burst_powerlaw}
} 

The value of the exponent $\tau=5/2$ is universal for a broad class of
failure threshold distributions 
\cite{kloster_burst_1997,pradhan_failure_2010,hidalgo_avalanche_2009}. 
\begin{figure}%[!h]
  \begin{center}
\epsfig{bbllx=20,bblly=10,bburx=360,bbury=300,file=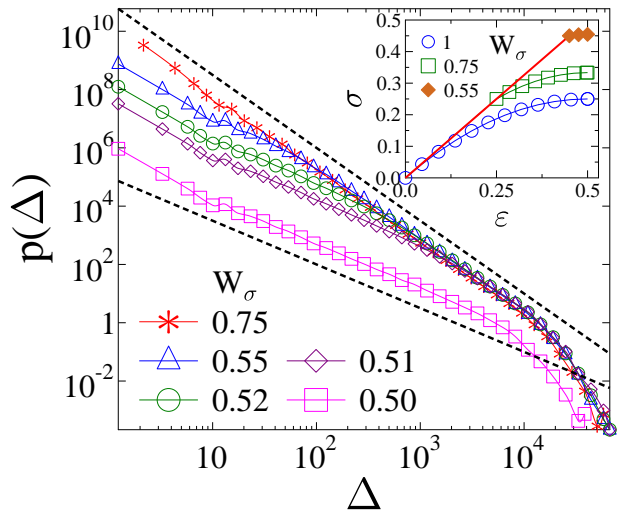, width=8.0cm}
   \caption{(Color online) Burst size distributions $p(\Delta)$ 
   for several values of strength disorder 
   $W_{\sigma}$ in the absence of stiffness disorder $W_{E}=0$.
Approaching the brittle regime $W_{\sigma} \to 1/2$ the well known result of the 
crossover in $\tau$ from $5/2$ to $3/2$ is recovered.
The two straight lines represent power laws with exponents $3/2$ and $5/2$.
The inset presents the constitutive behaviour of the system for three values of 
$W_{\sigma}$.
} \label{uniburst1}
\end{center} 
\end{figure}
Avalanches occur until the disorder
is high enough in the system $W_{\sigma}> \sigma_+/2$ 
\cite{pradhan_crossover_2005-1,raischel_local_2006}. In the limiting case 
$W_{\sigma} \rightarrow \sigma_+/2$ the quasi-brittle region where avalanche 
precursors occur, shrinks such that when
$W_{\sigma} \leq \sigma_+/2$ the response of the bundle becomes perfectly 
brittle
and the system collapses without having any finite size avalanches. 
The constitutive behaviour of the system under stress controlled 
loading is shown in the inset of Fig.\ \ref{uniburst1} 
for a few values of $W_{\sigma}$, where the shrinking of the quasi-brittle
regime can be observed.
Approaching the brittle limit the power law size distribution of avalanches 
prevails,
however, $p(\Delta)$ exhibits a crossover to a lower exponent $\tau=3/2$ in 
agreement with previous findings 
\cite{hidalgo_avalanche_2009,pradhan_crossover_2005-1,raischel_local_2006}.
This behavior can be seen in Fig.\ \ref{uniburst1} where avalanche size 
distributions 
$p(\Delta)$ of our model are presented for several values of $W_{\sigma}$ 
at zero stiffness disorder $W_E=0$. Approaching the brittle limit the lower 
value of the avalanche size exponent $\tau=3/2$ means that the fraction of 
large size events gets higher since the breaking 
of stronger fibers can trigger more secondary breakings.
Avalanches of breaking fibers are analogous to acoustic outbreaks generated by
the nucleation and propagation of cracks in heterogeneous solids under an 
increasing stress 
\cite{niccolini_acoustic_2011,salminen_acoustic_2002,diodati_acoustic_1991}. 
Recording the acoustic waves that emanate from the material is an indicator of
the breaking phenomena on the microscopic level. It has been addressed that the 
crossover in $\tau$ could be exploited to forecast the imminent catastrophic 
failure \cite{pradhan_precursors_2002,pradhan_crossover_2005-1}. 

\begin{figure}%[!h] 
  \begin{center}
\epsfig{bbllx=25,bblly=120,bburx=730,bbury=730,
file=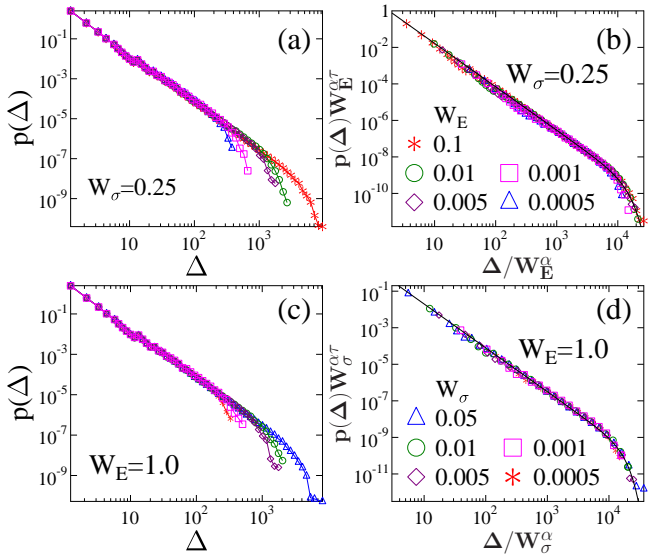,width=8.5cm}
  \caption{(Color online) $(a)$ Size distribution of bursts for $W_{\sigma}= 0.25$ varying the 
  stiffness disorder $W_E$ in a broad range. $(b)$ Scaling collapse of the distributions
  of $(a)$. Best collapse is achieved with the exponent $\alpha = 1/2$. 
  In $(a)$ and $(b)$ the same legend is used.
  $(c)$ Size distribution of bursts for $W_E=1$ varying the amount of strength disorder 
  $W_{\sigma}$ in a broad range. $(d)$ Scaling collapse of the distributions of $(c)$
  using the exponent $\alpha=1/2$. In $(c)$ and $(d)$ the same legend is used.
\label{fig:uniburst_scaling}}
\end{center} 
\end{figure}

In Sec.\ \ref{sec:macro} the analysis of the fraction of broken fibers $P_b$ at the peak load 
has already shown that adding a second source of disorder 
to an otherwise brittle system, quasi-brittle behavior emerges, i.e.\ $P_b$ is never 
zero when $W_{\sigma}, W_E>0$. It has the interesting consequence that macroscopic 
failure is always preceded by finite avalanches which behave as precursors to failure. 
To demonstrate this effect in Fig.\ \ref{fig:uniburst_scaling}$(a)$ avalanche size distributions 
are presented for $W_{\sigma}=0.25$ which should provide perfectly brittle behavior 
(no stable avalanches) if strength is the only source of disorder. 
Simulations revealed that for any finite value of $W_E$ the size distribution 
$p(\Delta)$ has a power law behavior as in Eq.\ (\ref{eq:burst_powerlaw}) 
followed by a stretched exponential cutoff.
For the value of the power
law exponent the usual mean field value $\tau=5/2$ was obtained.
It can be observed that the amount of disorder $W_E$ only controls the total 
number of avalanches and the cutoff burst size of the distributions. 
Figure \ref{fig:uniburst_scaling}$(b)$ shows that rescaling the burst size 
by the $\alpha=1/2$ power of $W_E$ all the distributions can be collapsed 
on a master curve. This high quality scaling collapse demonstrates that 
the exponent $\tau$ is independent of the amount disorder, even in the limit 
of very low $W_E$ the same exponent $\tau = 5/2$ is retained. Note that due to 
the normalization of the distributions along the vertical axis scaling is done
with the product of the two exponents $\alpha$ and $\tau$.

We have pointed out in Ref.\ \cite{0295-5075-95-1-16004} that 
when stiffness is the only source of disorder $W_{\sigma}=0$ perfectly brittle behavior 
occurs for any value of $W_E$. However, in the present study we have shown that 
adding strength disorder leads to stabilization. Avalanche size distributions 
are presented in Fig.\ \ref{fig:uniburst_scaling}$(c)$ for $W_E=1$ varying the amount of strength
disorder in a broad range. Again the same functional form 
occurs as in Fig.\ \ref{fig:uniburst_scaling}$(a)$  with the 
same exponent $\tau = 5/2$ for all the parameter sets.
The scaling collapse in Fig.\ \ref{fig:uniburst_scaling}$(d)$ is also obtained with 
the exponent $\alpha = 1/2$ as for the case of varying stiffness disorder.
The data collapse analysis 
also implies that the cutoff burst size $\Delta_c$
of the distributions has a power law dependence on the 
amount of disorder in the form $\Delta_c \sim W_E^{\alpha}$ and
$\Delta_c\sim W_{\sigma}^{\alpha}$ when the stiffness and strength disorder 
are varied, respectively, in the limit of low disorder.

A very important consequence of the above numerical analysis is that when approaching
the brittle limit $W_E\to 0$ for $W_{\sigma}<1/2$ and $W_{\sigma}\to 0$ for $W_E > 0$, 
the avalanche size distributions do not show any crossover to a lower exponent.
Reducing the amount of disorder both the number and size of avalanches decreases,
however, the value of the power law exponent $\tau$ remains robust.
Crossover occurs when solely strength disorder is present $W_E=0$.

Note that no similar scaling behaviour can be observed when only one source of disorder 
is present: For constant fiber strength, the system has a perfectly brittle behaviour
at any values of $W_E$. When strength is the only source of disorder the bundle 
becomes critical already when $W_{\sigma}$ approaches $1/2$. It has the consequence that 
in the limit $W_{\sigma} \to 1/2$ the characteristic avalanche size increases, i.e.\ 
just the opposite happens to what we have presented above. When the two disorders are blended
the reason of the decreasing avalanche activity is that for any $W_E$ and $W_{\sigma}$ 
the threshold distribution $h(\varepsilon_{th})$ starts from a zero value even if the thresholds 
have a finite lower bound $\varepsilon_{th}^{min}>0$.

\section{Discussion}
The fracture of disordered materials proceeds in bursts which can be recorded 
in the form of acoustic noise. Measuring crackling noise is the primary source
of information about the microscopic dynamics and time evolution of fracture.
From laboratory experiments through engineering constructions to
the scale of natural catastrophes forecasting techniques of imminent global
failure strongly rely on identifying signatures in the evolving temporal sequence
of breaking bursts \cite{sammonds_role_1992,PhysRevLett.112.115502}. 
When the disorder is absent or its amount is not sufficiently high,
failure occurs in a catastrophic manner without any precursors. Hence, enhancing 
the quasi-brittle nature of fracture by controlling disorder is of high practical importance. 

In the present paper we considered this problem in the framework of fiber
bundle models. This approach  provides a simple representation of the disorder 
and allows for the investigation of the microscopic dynamics of the failure process 
under various
types of loading conditions. The classical setup of FBMs assumes constant Young 
modulus of
fibers so that the heterogeneous microstructure is solely captured by the random 
strength of fibers. 
Here we proposed an extension of FBMs by considering simultaneously
two sources of disorder, i.e.\ both the strength and stiffness of fibers are 
random variables independent of each other.  We carried out a 
detailed analytical and numerical investigation of the fracture process of the system 
under quasi-statically increasing external load both on the macro and micro scales.

For the case of global load sharing we showed that introducing a second source of disorder 
stabilizes the system in the sense that a bundle which has a perfectly brittle
behavior becomes quasi-brittle whenever a finite amount of disorder of the 
other field is added. 
We gave a general analytical derivation of the constitutive equation and of 
the stability criterion of the system in terms of the disorder distributions.
For the purpose of numerical investigations an efficient simulation technique
was worked out.
As a specific case we considered uniformly distributed strength and stiffness
of fibers controlling the amount of disorder by the width of the distributions.
Investigating the microscopic process of failure 
we pointed out that the size of crackling bursts is power law distributed 
followed by an exponential cutoff. The power law exponent proved to be equal
to the usual mean field exponent $5/2$ without having any crossover to a lower
value when approaching the limit of perfect brittleness. The amount of disorder only
controls the number of avalanches and their cutoff size.
The origin of the stabilization mechanism is that the distribution of the relevant 
failure threshold, obtained as the convolution of the stiffness and strength 
distributions of fibers, starts from a zero value even if the thresholds 
have a finite lower bound. 

Recently, the problem of mixing strength and stiffness disorder has been considered
in a simplified modelling framework: in Ref.\ \cite{roy_recipe_2015} 
a bundle was composed of a few groups of fibers 
of different Young moduli having uniformly distributed failure strength.
Approximate calculations showed that increasing the number of groups
of equally spaced Young modulus values the bundle retains the quasi-brittle 
behaviour for narrower strength distributions. Our analytical results 
provide a general understanding of the findings of Ref.\ \cite{roy_recipe_2015} 
with the additional outcome that the continuous stiffness distribution 
of our study corresponds to an infinite number of groups of fibers where 
stability is ensured for any finite amount of strength disorder.

To test the generality of our results, we also considered the case where 
the breaking threshold and Young modulus of fibers follow a Weibull distribution 
with a lower cutoff $x_-$. Here $x$ stands for both strength $\sigma_{th}$ and 
stiffness $E$. Simulations performed with several values of $x_-$
verified that any finite amount of disorder leads to quasi-brittle behavior 
of the bundle when two sources of disorder are present. Furthermore, 
the burst size distribution exponent $\tau$ displayed a crossover from $5/2$ 
to $3/2$ only when the strength disorder was reduced in the absence of 
stiffness disorder in agreement with \cite{pradhan_crossover_2006}.

Besides their theoretical importance our results have practical relevance 
for materials' design, the controlled blending of stiffness and strength
disorder is a promissing way to increase the safety of constructions. 
Most of our results are formulated in a general way so that they can be applied 
to any strength and stiffness distributions used in engineering and materials science.
Biological materials exhibit a broader variety of stiffness and strength than engineering 
materials which could also be captured in the framework of our model.
An important limitation of our study that has to be resolved is the assumption
that thrength and stiffness of material elements are uncorrelated. In real materials
correlations naturally develop such that higher stiffness may be accompanied
by higher strength. Work is in progress to capture these types of correlation 
in our model.

\begin{acknowledgments}
We thank financial support of the projects TAMOP-4.2.2.A-11/1/KONV-2012-0036, 
and OTKA K84157.
\end{acknowledgments}

\bibliography{/home/feri/papers/statphys_fracture}

\begin{thebibliography}{34}%
\makeatletter
\providecommand \@ifxundefined [1]{%
 \@ifx{#1\undefined}
}%
\providecommand \@ifnum [1]{%
 \ifnum #1\expandafter \@firstoftwo
 \else \expandafter \@secondoftwo
 \fi
}%
\providecommand \@ifx [1]{%
 \ifx #1\expandafter \@firstoftwo
 \else \expandafter \@secondoftwo
 \fi
}%
\providecommand \natexlab [1]{#1}%
\providecommand \enquote  [1]{``#1''}%
\providecommand \bibnamefont  [1]{#1}%
\providecommand \bibfnamefont [1]{#1}%
\providecommand \citenamefont [1]{#1}%
\providecommand \href@noop [0]{\@secondoftwo}%
\providecommand \href [0]{\begingroup \@sanitize@url \@href}%
\providecommand \@href[1]{\@@startlink{#1}\@@href}%
\providecommand \@@href[1]{\endgroup#1\@@endlink}%
\providecommand \@sanitize@url [0]{\catcode `\\12\catcode `\$12\catcode
  `\&12\catcode `\#12\catcode `\^12\catcode `\_12\catcode `\%12\relax}%
\providecommand \@@startlink[1]{}%
\providecommand \@@endlink[0]{}%
\providecommand \url  [0]{\begingroup\@sanitize@url \@url }%
\providecommand \@url [1]{\endgroup\@href {#1}{\urlprefix }}%
\providecommand \urlprefix  [0]{URL }%
\providecommand \Eprint [0]{\href }%
\providecommand \doibase [0]{http://dx.doi.org/}%
\providecommand \selectlanguage [0]{\@gobble}%
\providecommand \bibinfo  [0]{\@secondoftwo}%
\providecommand \bibfield  [0]{\@secondoftwo}%
\providecommand \translation [1]{[#1]}%
\providecommand \BibitemOpen [0]{}%
\providecommand \bibitemStop [0]{}%
\providecommand \bibitemNoStop [0]{.\EOS\space}%
\providecommand \EOS [0]{\spacefactor3000\relax}%
\providecommand \BibitemShut  [1]{\csname bibitem#1\endcsname}%
\let\auto@bib@innerbib\@empty
%</preamble>
\bibitem [{\citenamefont {Herrmann}\ and\ \citenamefont
  {Roux}(1990)}]{herrmann_statistical_1990}%
  \BibitemOpen
  \bibinfo {editor} {\bibfnamefont {H.~J.}\ \bibnamefont {Herrmann}}\ and\
  \bibinfo {editor} {\bibfnamefont {S.}~\bibnamefont {Roux}},\ eds.,\
  \href@noop {} {\emph {\bibinfo {title} {Statistical models for the fracture
  of disordered media}}},\ Random materials and processes\ (\bibinfo
  {publisher} {Elsevier},\ \bibinfo {address} {Amsterdam},\ \bibinfo {year}
  {1990})\BibitemShut {NoStop}%
\bibitem [{\citenamefont {Alava}\ \emph {et~al.}(2006)\citenamefont {Alava},
  \citenamefont {Nukala},\ and\ \citenamefont
  {Zapperi}}]{alava_statistical_2006}%
  \BibitemOpen
  \bibfield  {author} {\bibinfo {author} {\bibfnamefont {M.}~\bibnamefont
  {Alava}}, \bibinfo {author} {\bibfnamefont {P.~K.}\ \bibnamefont {Nukala}}, \
  and\ \bibinfo {author} {\bibfnamefont {S.}~\bibnamefont {Zapperi}},\
  }\href@noop {} {\bibfield  {journal} {\bibinfo  {journal} {Adv. Phys.}\
  }\textbf {\bibinfo {volume} {55}},\ \bibinfo {pages} {349–476} (\bibinfo
  {year} {2006})}\BibitemShut {NoStop}%
\bibitem [{\citenamefont {Biswas}\ \emph {et~al.}(2015)\citenamefont {Biswas},
  \citenamefont {Ray},\ and\ \citenamefont
  {Chakrabarti}}]{book_chakrabarti_2015}%
  \BibitemOpen
  \bibfield  {author} {\bibinfo {author} {\bibfnamefont {S.}~\bibnamefont
  {Biswas}}, \bibinfo {author} {\bibfnamefont {P.}~\bibnamefont {Ray}}, \ and\
  \bibinfo {author} {\bibfnamefont {B.~K.}\ \bibnamefont {Chakrabarti}},\
  }\href@noop {} {\emph {\bibinfo {title} {Statistical Physics of Fracture,
  Beakdown, and Earthquake: Effects of Disorder and Heterogeneity}}},\
  Statistical Physics of Fracture and Breakdown\ (\bibinfo  {publisher} {John
  Wiley \& Sons},\ \bibinfo {address} {New York},\ \bibinfo {year}
  {2015})\BibitemShut {NoStop}%
\bibitem [{\citenamefont {Ramos}\ \emph {et~al.}(2013)\citenamefont {Ramos},
  \citenamefont {Cortet}, \citenamefont {Ciliberto},\ and\ \citenamefont
  {Vanel}}]{PhysRevLett.110.165506}%
  \BibitemOpen
  \bibfield  {author} {\bibinfo {author} {\bibfnamefont {O.}~\bibnamefont
  {Ramos}}, \bibinfo {author} {\bibfnamefont {P.-P.}\ \bibnamefont {Cortet}},
  \bibinfo {author} {\bibfnamefont {S.}~\bibnamefont {Ciliberto}}, \ and\
  \bibinfo {author} {\bibfnamefont {L.}~\bibnamefont {Vanel}},\ }\href@noop {}
  {\bibfield  {journal} {\bibinfo  {journal} {Phys. Rev. Lett.}\ }\textbf
  {\bibinfo {volume} {110}},\ \bibinfo {pages} {165506} (\bibinfo {year}
  {2013})}\BibitemShut {NoStop}%
\bibitem [{\citenamefont {Hal\'asz}\ \emph {et~al.}(2012)\citenamefont
  {Hal\'asz}, \citenamefont {Danku},\ and\ \citenamefont
  {Kun}}]{PhysRevE.85.016116}%
  \BibitemOpen
  \bibfield  {author} {\bibinfo {author} {\bibfnamefont {Z.}~\bibnamefont
  {Hal\'asz}}, \bibinfo {author} {\bibfnamefont {Z.}~\bibnamefont {Danku}}, \
  and\ \bibinfo {author} {\bibfnamefont {F.}~\bibnamefont {Kun}},\ }\href@noop
  {} {\bibfield  {journal} {\bibinfo  {journal} {Phys. Rev. E}\ }\textbf
  {\bibinfo {volume} {85}},\ \bibinfo {pages} {016116} (\bibinfo {year}
  {2012})}\BibitemShut {NoStop}%
\bibitem [{\citenamefont {Vasseur}\ \emph {et~al.}(2015)\citenamefont
  {Vasseur}, \citenamefont {Wadsworth}, \citenamefont {Lavallée},
  \citenamefont {Bell}, \citenamefont {Main},\ and\ \citenamefont
  {Dingwell}}]{vasseur_heterogeneity:_2015}%
  \BibitemOpen
  \bibfield  {author} {\bibinfo {author} {\bibfnamefont {J.}~\bibnamefont
  {Vasseur}}, \bibinfo {author} {\bibfnamefont {F.~B.}\ \bibnamefont
  {Wadsworth}}, \bibinfo {author} {\bibfnamefont {Y.}~\bibnamefont
  {Lavallée}}, \bibinfo {author} {\bibfnamefont {A.~F.}\ \bibnamefont {Bell}},
  \bibinfo {author} {\bibfnamefont {I.~G.}\ \bibnamefont {Main}}, \ and\
  \bibinfo {author} {\bibfnamefont {D.~B.}\ \bibnamefont {Dingwell}},\ }\href
  {\doibase 10.1038/srep13259} {\bibfield  {journal} {\bibinfo  {journal} {Sci.
  Rep.}\ }\textbf {\bibinfo {volume} {5}},\ \bibinfo {pages} {13259} (\bibinfo
  {year} {2015})}\BibitemShut {NoStop}%
\bibitem [{\citenamefont {Kun}\ and\ \citenamefont
  {Herrmann}(2000)}]{kun_damage_2000}%
  \BibitemOpen
  \bibfield  {author} {\bibinfo {author} {\bibfnamefont {F.}~\bibnamefont
  {Kun}}\ and\ \bibinfo {author} {\bibfnamefont {H.~J.}\ \bibnamefont
  {Herrmann}},\ }\href@noop {} {\bibfield  {journal} {\bibinfo  {journal} {J.
  Mat. Sci.}\ }\textbf {\bibinfo {volume} {35}},\ \bibinfo {pages} {4685}
  (\bibinfo {year} {2000})}\BibitemShut {NoStop}%
\bibitem [{\citenamefont {Hidalgo}\ \emph {et~al.}(2008)\citenamefont
  {Hidalgo}, \citenamefont {Kov\'acs}, \citenamefont {Pagonabarraga},\ and\
  \citenamefont {Kun}}]{hidalgo_universality_2008}%
  \BibitemOpen
  \bibfield  {author} {\bibinfo {author} {\bibfnamefont {R.~C.}\ \bibnamefont
  {Hidalgo}}, \bibinfo {author} {\bibfnamefont {K.}~\bibnamefont {Kov\'acs}},
  \bibinfo {author} {\bibfnamefont {I.}~\bibnamefont {Pagonabarraga}}, \ and\
  \bibinfo {author} {\bibfnamefont {F.}~\bibnamefont {Kun}},\ }\href@noop {}
  {\bibfield  {journal} {\bibinfo  {journal} {Europhys. Lett.}\ }\textbf
  {\bibinfo {volume} {81}},\ \bibinfo {pages} {54005} (\bibinfo {year}
  {2008})}\BibitemShut {NoStop}%
\bibitem [{\citenamefont {Hidalgo}\ \emph {et~al.}(2009)\citenamefont
  {Hidalgo}, \citenamefont {Kun}, \citenamefont {Kov\'acs},\ and\ \citenamefont
  {Pagonabarraga}}]{hidalgo_avalanche_2009}%
  \BibitemOpen
  \bibfield  {author} {\bibinfo {author} {\bibfnamefont {R.~C.}\ \bibnamefont
  {Hidalgo}}, \bibinfo {author} {\bibfnamefont {F.}~\bibnamefont {Kun}},
  \bibinfo {author} {\bibfnamefont {K.}~\bibnamefont {Kov\'acs}}, \ and\
  \bibinfo {author} {\bibfnamefont {I.}~\bibnamefont {Pagonabarraga}},\
  }\href@noop {} {\bibfield  {journal} {\bibinfo  {journal} {Phys. Rev. E}\
  }\textbf {\bibinfo {volume} {80}},\ \bibinfo {pages} {051108} (\bibinfo
  {year} {2009})}\BibitemShut {NoStop}%
\bibitem [{\citenamefont {Pradhan}\ \emph {et~al.}(2010)\citenamefont
  {Pradhan}, \citenamefont {Hansen},\ and\ \citenamefont
  {Chakrabarti}}]{pradhan_failure_2010}%
  \BibitemOpen
  \bibfield  {author} {\bibinfo {author} {\bibfnamefont {S.}~\bibnamefont
  {Pradhan}}, \bibinfo {author} {\bibfnamefont {A.}~\bibnamefont {Hansen}}, \
  and\ \bibinfo {author} {\bibfnamefont {B.~K.}\ \bibnamefont {Chakrabarti}},\
  }\href@noop {} {\bibfield  {journal} {\bibinfo  {journal} {Rev. Mod. Phys.}\
  }\textbf {\bibinfo {volume} {82}},\ \bibinfo {pages} {499} (\bibinfo {year}
  {2010})}\BibitemShut {NoStop}%
\bibitem [{\citenamefont {Nukala}\ \emph {et~al.}(2004)\citenamefont {Nukala},
  \citenamefont {Simunovic},\ and\ \citenamefont
  {Zapperi}}]{nukala_percolation_2004}%
  \BibitemOpen
  \bibfield  {author} {\bibinfo {author} {\bibfnamefont {P.~V.~V.}\
  \bibnamefont {Nukala}}, \bibinfo {author} {\bibfnamefont {S.}~\bibnamefont
  {Simunovic}}, \ and\ \bibinfo {author} {\bibfnamefont {S.}~\bibnamefont
  {Zapperi}},\ }\href@noop {} {\bibfield  {journal} {\bibinfo  {journal} {J.
  Stat. Mech: Theor. Exp.}\ ,\ \bibinfo {pages} {P08001}} (\bibinfo {year}
  {2004})}\BibitemShut {NoStop}%
\bibitem [{\citenamefont {{D'Addetta}}\ \emph {et~al.}(2002)\citenamefont
  {{D'Addetta}}, \citenamefont {Kun},\ and\ \citenamefont
  {Ramm}}]{daddetta_application_2002}%
  \BibitemOpen
  \bibfield  {author} {\bibinfo {author} {\bibfnamefont {G.~A.}\ \bibnamefont
  {{D'Addetta}}}, \bibinfo {author} {\bibfnamefont {F.}~\bibnamefont {Kun}}, \
  and\ \bibinfo {author} {\bibfnamefont {E.}~\bibnamefont {Ramm}},\ }\href@noop
  {} {\bibfield  {journal} {\bibinfo  {journal} {Gran. Matt.}\ }\textbf
  {\bibinfo {volume} {4}},\ \bibinfo {pages} {77} (\bibinfo {year}
  {2002})}\BibitemShut {NoStop}%
\bibitem [{\citenamefont {Carmona}\ \emph {et~al.}(2008)\citenamefont
  {Carmona}, \citenamefont {Wittel}, \citenamefont {Kun},\ and\ \citenamefont
  {Herrmann}}]{carmona_fragmentation_2008}%
  \BibitemOpen
  \bibfield  {author} {\bibinfo {author} {\bibfnamefont {H.~A.}\ \bibnamefont
  {Carmona}}, \bibinfo {author} {\bibfnamefont {F.~K.}\ \bibnamefont {Wittel}},
  \bibinfo {author} {\bibfnamefont {F.}~\bibnamefont {Kun}}, \ and\ \bibinfo
  {author} {\bibfnamefont {H.~J.}\ \bibnamefont {Herrmann}},\ }\href@noop {}
  {\bibfield  {journal} {\bibinfo  {journal} {Phys. Rev. E}\ }\textbf {\bibinfo
  {volume} {77}},\ \bibinfo {pages} {051302} (\bibinfo {year}
  {2008})}\BibitemShut {NoStop}%
\bibitem [{\citenamefont {Ergenzinger}\ \emph {et~al.}(2010)\citenamefont
  {Ergenzinger}, \citenamefont {Seifried},\ and\ \citenamefont
  {Eberhard}}]{ergenzinger_discrete_2010}%
  \BibitemOpen
  \bibfield  {author} {\bibinfo {author} {\bibfnamefont {C.}~\bibnamefont
  {Ergenzinger}}, \bibinfo {author} {\bibfnamefont {R.}~\bibnamefont
  {Seifried}}, \ and\ \bibinfo {author} {\bibfnamefont {P.}~\bibnamefont
  {Eberhard}},\ }\href@noop {} {\bibfield  {journal} {\bibinfo  {journal}
  {Gran. Matt.}\ }\textbf {\bibinfo {volume} {13}},\ \bibinfo {pages} {341}
  (\bibinfo {year} {2010})}\BibitemShut {NoStop}%
\bibitem [{\citenamefont {Okumura}(2004)}]{okumura_toughness_2004}%
  \BibitemOpen
  \bibfield  {author} {\bibinfo {author} {\bibfnamefont {K.}~\bibnamefont
  {Okumura}},\ }\href@noop {} {\bibfield  {journal} {\bibinfo  {journal}
  {Europhys. Lett.}\ }\textbf {\bibinfo {volume} {67}},\ \bibinfo {pages} {470}
  (\bibinfo {year} {2004})}\BibitemShut {NoStop}%
\bibitem [{\citenamefont {Urabe}\ and\ \citenamefont
  {Takesue}(2010)}]{urabe_fracture_2010}%
  \BibitemOpen
  \bibfield  {author} {\bibinfo {author} {\bibfnamefont {C.}~\bibnamefont
  {Urabe}}\ and\ \bibinfo {author} {\bibfnamefont {S.}~\bibnamefont
  {Takesue}},\ }\href@noop {} {\bibfield  {journal} {\bibinfo  {journal} {Phys.
  Rev. E}\ }\textbf {\bibinfo {volume} {82}},\ \bibinfo {pages} {016106}
  (\bibinfo {year} {2010})}\BibitemShut {NoStop}%
\bibitem [{\citenamefont {Okumura}\ and\ \citenamefont
  {de~Gennes}(2001)}]{okumura_why_2001}%
  \BibitemOpen
  \bibfield  {author} {\bibinfo {author} {\bibfnamefont {K.}~\bibnamefont
  {Okumura}}\ and\ \bibinfo {author} {\bibfnamefont {P.}~\bibnamefont
  {de~Gennes}},\ }\href@noop {} {\bibfield  {journal} {\bibinfo  {journal}
  {Eur. Phys. Jour. E}\ }\textbf {\bibinfo {volume} {4}},\ \bibinfo {pages}
  {121} (\bibinfo {year} {2001})}\BibitemShut {NoStop}%
\bibitem [{\citenamefont {Nukala}\ and\ \citenamefont
  {Simunovic}(2005)}]{nukala_continuous_2005}%
  \BibitemOpen
  \bibfield  {author} {\bibinfo {author} {\bibfnamefont {P.~K.}\ \bibnamefont
  {Nukala}}\ and\ \bibinfo {author} {\bibfnamefont {S.}~\bibnamefont
  {Simunovic}},\ }\href@noop {} {\bibfield  {journal} {\bibinfo  {journal}
  {Biomaterials}\ }\textbf {\bibinfo {volume} {26}},\ \bibinfo {pages} {6087}
  (\bibinfo {year} {2005})}\BibitemShut {NoStop}%
\bibitem [{\citenamefont {Layton}\ and\ \citenamefont
  {Sastry}(2006)}]{layton_equal_2006}%
  \BibitemOpen
  \bibfield  {author} {\bibinfo {author} {\bibfnamefont {B.~E.}\ \bibnamefont
  {Layton}}\ and\ \bibinfo {author} {\bibfnamefont {A.~M.}\ \bibnamefont
  {Sastry}},\ }\href@noop {} {\bibfield  {journal} {\bibinfo  {journal} {Acta
  Biomaterialia}\ }\textbf {\bibinfo {volume} {68}},\ \bibinfo {pages} {612}
  (\bibinfo {year} {2006})}\BibitemShut {NoStop}%
\bibitem [{\citenamefont {Barthelat}(2007)}]{barthelat_biomimetics_2007}%
  \BibitemOpen
  \bibfield  {author} {\bibinfo {author} {\bibfnamefont {F.}~\bibnamefont
  {Barthelat}},\ }\href@noop {} {\bibfield  {journal} {\bibinfo  {journal}
  {Phil. Trans. Roy. Soc. A: Math. Phys. Eng. Sci.}\ }\textbf {\bibinfo
  {volume} {365}},\ \bibinfo {pages} {2907 } (\bibinfo {year}
  {2007})}\BibitemShut {NoStop}%
\bibitem [{\citenamefont {Yoshioka}\ \emph {et~al.}(2008)\citenamefont
  {Yoshioka}, \citenamefont {Kun},\ and\ \citenamefont
  {Ito}}]{yoshioka_size_2008}%
  \BibitemOpen
  \bibfield  {author} {\bibinfo {author} {\bibfnamefont {N.}~\bibnamefont
  {Yoshioka}}, \bibinfo {author} {\bibfnamefont {F.}~\bibnamefont {Kun}}, \
  and\ \bibinfo {author} {\bibfnamefont {N.}~\bibnamefont {Ito}},\ }\href@noop
  {} {\bibfield  {journal} {\bibinfo  {journal} {Phys. Rev. Lett.}\ }\textbf
  {\bibinfo {volume} {101}},\ \bibinfo {pages} {145502} (\bibinfo {year}
  {2008})}\BibitemShut {NoStop}%
\bibitem [{\citenamefont {Karpas}\ and\ \citenamefont
  {Kun}(2011)}]{0295-5075-95-1-16004}%
  \BibitemOpen
  \bibfield  {author} {\bibinfo {author} {\bibfnamefont {E.}~\bibnamefont
  {Karpas}}\ and\ \bibinfo {author} {\bibfnamefont {F.}~\bibnamefont {Kun}},\
  }\href@noop {} {\bibfield  {journal} {\bibinfo  {journal} {Europhys. Lett.}\
  }\textbf {\bibinfo {volume} {95}},\ \bibinfo {pages} {16004} (\bibinfo {year}
  {2011})}\BibitemShut {NoStop}%
\bibitem [{\citenamefont {Pradhan}\ \emph {et~al.}(2005)\citenamefont
  {Pradhan}, \citenamefont {Hansen},\ and\ \citenamefont
  {Hemmer}}]{pradhan_crossover_2005-1}%
  \BibitemOpen
  \bibfield  {author} {\bibinfo {author} {\bibfnamefont {S.}~\bibnamefont
  {Pradhan}}, \bibinfo {author} {\bibfnamefont {A.}~\bibnamefont {Hansen}}, \
  and\ \bibinfo {author} {\bibfnamefont {P.~C.}\ \bibnamefont {Hemmer}},\
  }\href@noop {} {\bibfield  {journal} {\bibinfo  {journal} {Phys. Rev. Lett.}\
  }\textbf {\bibinfo {volume} {95}},\ \bibinfo {pages} {125501} (\bibinfo
  {year} {2005})}\BibitemShut {NoStop}%
\bibitem [{\citenamefont {Raischel}\ \emph {et~al.}(2006)\citenamefont
  {Raischel}, \citenamefont {Kun},\ and\ \citenamefont
  {Herrmann}}]{raischel_local_2006}%
  \BibitemOpen
  \bibfield  {author} {\bibinfo {author} {\bibfnamefont {F.}~\bibnamefont
  {Raischel}}, \bibinfo {author} {\bibfnamefont {F.}~\bibnamefont {Kun}}, \
  and\ \bibinfo {author} {\bibfnamefont {H.~J.}\ \bibnamefont {Herrmann}},\
  }\href@noop {} {\bibfield  {journal} {\bibinfo  {journal} {Phys. Rev. E}\
  }\textbf {\bibinfo {volume} {74}},\ \bibinfo {pages} {035104} (\bibinfo
  {year} {2006})}\BibitemShut {NoStop}%
\bibitem [{\citenamefont {Kloster}\ \emph {et~al.}(1997)\citenamefont
  {Kloster}, \citenamefont {Hansen},\ and\ \citenamefont
  {Hemmer}}]{kloster_burst_1997}%
  \BibitemOpen
  \bibfield  {author} {\bibinfo {author} {\bibfnamefont {M.}~\bibnamefont
  {Kloster}}, \bibinfo {author} {\bibfnamefont {A.}~\bibnamefont {Hansen}}, \
  and\ \bibinfo {author} {\bibfnamefont {P.~C.}\ \bibnamefont {Hemmer}},\
  }\href@noop {} {\bibfield  {journal} {\bibinfo  {journal} {Phys. Rev. E}\
  }\textbf {\bibinfo {volume} {56}},\ \bibinfo {pages} {2615–2625} (\bibinfo
  {year} {1997})}\BibitemShut {NoStop}%
\bibitem [{\citenamefont {Kun}\ \emph {et~al.}(2000)\citenamefont {Kun},
  \citenamefont {Zapperi},\ and\ \citenamefont {Herrmann}}]{kun_damage_2000-1}%
  \BibitemOpen
  \bibfield  {author} {\bibinfo {author} {\bibfnamefont {F.}~\bibnamefont
  {Kun}}, \bibinfo {author} {\bibfnamefont {S.}~\bibnamefont {Zapperi}}, \ and\
  \bibinfo {author} {\bibfnamefont {H.~J.}\ \bibnamefont {Herrmann}},\
  }\href@noop {} {\bibfield  {journal} {\bibinfo  {journal} {Eur. Phys. J. B}\
  }\textbf {\bibinfo {volume} {17}},\ \bibinfo {pages} {269} (\bibinfo {year}
  {2000})}\BibitemShut {NoStop}%
\bibitem [{\citenamefont {Niccolini}\ \emph {et~al.}(2011)\citenamefont
  {Niccolini}, \citenamefont {Carpinteri}, \citenamefont {Lacidogna},\ and\
  \citenamefont {Manuello}}]{niccolini_acoustic_2011}%
  \BibitemOpen
  \bibfield  {author} {\bibinfo {author} {\bibfnamefont {G.}~\bibnamefont
  {Niccolini}}, \bibinfo {author} {\bibfnamefont {A.}~\bibnamefont
  {Carpinteri}}, \bibinfo {author} {\bibfnamefont {G.}~\bibnamefont
  {Lacidogna}}, \ and\ \bibinfo {author} {\bibfnamefont {A.}~\bibnamefont
  {Manuello}},\ }\href@noop {} {\bibfield  {journal} {\bibinfo  {journal}
  {Phys. Rev. Lett.}\ }\textbf {\bibinfo {volume} {106}},\ \bibinfo {pages}
  {108503} (\bibinfo {year} {2011})}\BibitemShut {NoStop}%
\bibitem [{\citenamefont {Salminen}\ \emph {et~al.}(2002)\citenamefont
  {Salminen}, \citenamefont {Tolvanen},\ and\ \citenamefont
  {Alava}}]{salminen_acoustic_2002}%
  \BibitemOpen
  \bibfield  {author} {\bibinfo {author} {\bibfnamefont {L.~I.}\ \bibnamefont
  {Salminen}}, \bibinfo {author} {\bibfnamefont {A.~I.}\ \bibnamefont
  {Tolvanen}}, \ and\ \bibinfo {author} {\bibfnamefont {M.~J.}\ \bibnamefont
  {Alava}},\ }\href@noop {} {\bibfield  {journal} {\bibinfo  {journal} {Phys.
  Rev. Lett.}\ }\textbf {\bibinfo {volume} {89}},\ \bibinfo {pages} {185503}
  (\bibinfo {year} {2002})}\BibitemShut {NoStop}%
\bibitem [{\citenamefont {Diodati}\ \emph {et~al.}(1991)\citenamefont
  {Diodati}, \citenamefont {Marchesoni},\ and\ \citenamefont
  {Piazza}}]{diodati_acoustic_1991}%
  \BibitemOpen
  \bibfield  {author} {\bibinfo {author} {\bibfnamefont {P.}~\bibnamefont
  {Diodati}}, \bibinfo {author} {\bibfnamefont {F.}~\bibnamefont {Marchesoni}},
  \ and\ \bibinfo {author} {\bibfnamefont {S.}~\bibnamefont {Piazza}},\
  }\href@noop {} {\bibfield  {journal} {\bibinfo  {journal} {Phys. Rev. Lett.}\
  }\textbf {\bibinfo {volume} {67}},\ \bibinfo {pages} {2239} (\bibinfo {year}
  {1991})}\BibitemShut {NoStop}%
\bibitem [{\citenamefont {Pradhan}\ and\ \citenamefont
  {Chakrabarti}(2002)}]{pradhan_precursors_2002}%
  \BibitemOpen
  \bibfield  {author} {\bibinfo {author} {\bibfnamefont {S.}~\bibnamefont
  {Pradhan}}\ and\ \bibinfo {author} {\bibfnamefont {B.~K.}\ \bibnamefont
  {Chakrabarti}},\ }\href@noop {} {\bibfield  {journal} {\bibinfo  {journal}
  {Phys. Rev. E}\ }\textbf {\bibinfo {volume} {65}},\ \bibinfo {pages} {016113}
  (\bibinfo {year} {2002})}\BibitemShut {NoStop}%
\bibitem [{\citenamefont {Sammonds}\ \emph {et~al.}(1992)\citenamefont
  {Sammonds}, \citenamefont {Meredith},\ and\ \citenamefont
  {Main}}]{sammonds_role_1992}%
  \BibitemOpen
  \bibfield  {author} {\bibinfo {author} {\bibfnamefont {P.~R.}\ \bibnamefont
  {Sammonds}}, \bibinfo {author} {\bibfnamefont {P.~G.}\ \bibnamefont
  {Meredith}}, \ and\ \bibinfo {author} {\bibfnamefont {I.~G.}\ \bibnamefont
  {Main}},\ }\href@noop {} {\bibfield  {journal} {\bibinfo  {journal} {Nature}\
  }\textbf {\bibinfo {volume} {359}},\ \bibinfo {pages} {228} (\bibinfo {year}
  {1992})}\BibitemShut {NoStop}%
\bibitem [{\citenamefont {Stojanova}\ \emph {et~al.}(2014)\citenamefont
  {Stojanova}, \citenamefont {Santucci}, \citenamefont {Vanel},\ and\
  \citenamefont {Ramos}}]{PhysRevLett.112.115502}%
  \BibitemOpen
  \bibfield  {author} {\bibinfo {author} {\bibfnamefont {M.}~\bibnamefont
  {Stojanova}}, \bibinfo {author} {\bibfnamefont {S.}~\bibnamefont {Santucci}},
  \bibinfo {author} {\bibfnamefont {L.}~\bibnamefont {Vanel}}, \ and\ \bibinfo
  {author} {\bibfnamefont {O.}~\bibnamefont {Ramos}},\ }\href@noop {}
  {\bibfield  {journal} {\bibinfo  {journal} {Phys. Rev. Lett.}\ }\textbf
  {\bibinfo {volume} {112}},\ \bibinfo {pages} {115502} (\bibinfo {year}
  {2014})}\BibitemShut {NoStop}%
\bibitem [{\citenamefont {Roy}\ and\ \citenamefont
  {Goswami}(2015)}]{roy_recipe_2015}%
  \BibitemOpen
  \bibfield  {author} {\bibinfo {author} {\bibfnamefont {S.}~\bibnamefont
  {Roy}}\ and\ \bibinfo {author} {\bibfnamefont {S.}~\bibnamefont {Goswami}},\
  }\href {http://arxiv.org/abs/1510.00687} {\bibfield  {journal} {\bibinfo
  {journal} {arXiv:1510.00687}\ } (\bibinfo {year} {2015})}\BibitemShut
  {NoStop}%
\bibitem [{\citenamefont {Pradhan}\ \emph {et~al.}(2006)\citenamefont
  {Pradhan}, \citenamefont {Hansen},\ and\ \citenamefont
  {Hemmer}}]{pradhan_crossover_2006}%
  \BibitemOpen
  \bibfield  {author} {\bibinfo {author} {\bibfnamefont {S.}~\bibnamefont
  {Pradhan}}, \bibinfo {author} {\bibfnamefont {A.}~\bibnamefont {Hansen}}, \
  and\ \bibinfo {author} {\bibfnamefont {P.~C.}\ \bibnamefont {Hemmer}},\
  }\href@noop {} {\bibfield  {journal} {\bibinfo  {journal} {Phys. Rev. E}\
  }\textbf {\bibinfo {volume} {74}},\ \bibinfo {pages} {016122} (\bibinfo
  {year} {2006})}\BibitemShut {NoStop}%
\end{thebibliography}%
\end{document}